\documentclass[aps,prb,twocolumn,superscriptaddress,footinbib]{revtex4-2}

\usepackage{graphicx}  
\usepackage{titletoc}
\usepackage{subfigure}
\usepackage{dcolumn} 
\usepackage{bm}
\usepackage{amssymb}   
\usepackage{comment}
\usepackage[colorlinks,linkcolor=blue,anchorcolor=blue,citecolor=blue,urlcolor=blue]{hyperref}
\usepackage{xcolor}
\usepackage{tabularx,booktabs,ragged2e,multirow}
\newcolumntype{C}{>{\centering\arraybackslash}X}
\usepackage{multirow}
\usepackage{amsmath}
\usepackage{mathrsfs}
\usepackage{mathcomp}
\usepackage{textcomp}
\usepackage{dsfont}
\usepackage{esint}
\usepackage{braket}
\usepackage{textgreek}
\usepackage{lipsum}
\usepackage{marvosym} 
\usepackage{centernot}
\usepackage{cancel}
\usepackage{dashrule}

\DeclareMathOperator{\sech}{sech}

\begin{document}
\preprint{APS/123-QED}

\title{Ultrafast excitation and topological soliton formation in incommensurate charge density wave states}

\author{Xiao-Xiao Zhang}
\affiliation{Department of Physics and Astronomy \& Stewart Blusson Quantum Matter Institute, University of British Columbia, Vancouver, BC, V6T 1Z4 Canada}
\author{Dirk Manske}
\affiliation{Max Planck Institute for Solid State Research, 70569 Stuttgart, Germany}
\author{Naoto Nagaosa}
\affiliation{Department of Applied Physics, University of Tokyo, Tokyo 113-8656, Japan}
\affiliation{RIKEN Center for Emergent Matter Science (CEMS), Wako, Saitama 351-0198, Japan}


\newcommand\dd{\mathrm{d}}
\newcommand\ii{\mathrm{i}}
\newcommand\ee{\mathrm{e}}
\newcommand\zz{\mathtt{z}}
\makeatletter
\let\newtitle\@title
\let\newauthor\@author
\def\ExtendSymbol#1#2#3#4#5{\ext@arrow 0099{\arrowfill@#1#2#3}{#4}{#5}}
\newcommand\LongEqual[2][]{\ExtendSymbol{=}{=}{=}{#1}{#2}}
\newcommand\LongArrow[2][]{\ExtendSymbol{-}{-}{\rightarrow}{#1}{#2}}
\newcommand{\cev}[1]{\reflectbox{\ensuremath{\vec{\reflectbox{\ensuremath{#1}}}}}}
\newcommand{\red}[1]{\textcolor{red}{#1}} 
\newcommand{\blue}[1]{\textcolor{blue}{#1}} 
\newcommand{\green}[1]{\textcolor{black}{#1}} 
\newcommand{\mycomment}[1]{} 
\makeatother

\begin{abstract}
Topological soliton is a nonperturbative excitation in commensurate density wave states 
and connects degenerate ground states. In incommensurate density wave states, ground states are continuously degenerate and topological soliton is reckoned to be smoothly connected to the perturbative phason excitation.
We study the ultrafast nonequilibrium dynamics due to photoexcited electron-hole pair in a one-dimensional chain with an incommensurate charge density wave ground state. Time-resolved 
evolution reveals both perturbative excitation of collective modes and nonperturbative topological phase transition due to creating novel topological solitons, where the continuous complex order parameter with amplitude and phase is essential. We identify the nontrivial phase-winding solitons in the complex plane unique to this nonequilibrium state and capture it by a low-energy effective model.
The perturbative temporal gap oscillation and the solitonic in-gap states enter the optical conductivity absorption edge and the spectral density related to spectroscopic measurement, providing concrete connections to real experiments.
\end{abstract}

\maketitle

\let\oldaddcontentsline\addcontentsline
\renewcommand{\addcontentsline}[3]{}


\section{Introduction}
Topological soliton is a subject of intense interest, governing various types of novel phenomena such as charge fractionalization and spin-charge separation\cite{Jackiw1976,Rajaraman}. 
It is regarded as 
the excited state connecting degenerate ground states, e.g., two different phases of bond dimerization in the representative half-filled polyacetylene where it can be detected by induced optical absorption\cite{Su1979,*Heeger1988}. Commensurate density waves in general have the similar situation with discrete multiple ground states and associated solitonic structures. In the incommensurate charge-density-wave (ICDW) state, however, the broken translational symmetry renders the ground states continuously degenerate: the phase Goldstone boson constitutes the low-lying excitations\cite{Lee1974,Brazovskii1976\mycomment{,Beekman2019}}, and soliton is usually not well-defined\cite{Gruener2018}.

On the other hand, significant progress in the precise access to ultrafast light-induced 
dynamics\cite{Giannetti2016,Nicoletti2016,Ostroverkhova2016,Collet2003,Koshihara2006,Loth2010,Udina2019} -- especially time-resolved spectroscopic techniques including angle-resolved photoemission spectroscopy (ARPES), optical bulk reflection or thin-film transmission measurements, 
electron diffuse scattering\cite{Stern2018,Cotret2019}, 
terahertz pump-probe scanning-tunneling microscopy (STM)\cite{Cocker2013,Eisele2014}, resonant inelastic X-ray scattering\cite{Cao2019,Mitrano2020}, 
etc. -- brings to the forefront nontrivial systems with broken symmetries via the nonlinear phenomena naturally arising 
from intense pulses and the subsequent nonequilibrium evolution\mycomment{ to be probed}. An intriguing question arises: What happens to the ICDW and if any soliton in the nonequilibrium?
A closely related and immediate need to the community is elucidating the real-time dynamics and possible nonlinear 
responses against optical disturbance in some prototypical low-dimensional quantum systems 
with especially broken continuous symmetries and ordered phases. 
This is in contrast to the field of Floquet engineering of topological phases developed in the past decade, where the photon's periodic driving influence is taken into account in another limit of steady states\cite{Oka2019,*Harper2020}.

To address these and motivated by the recent experimental interest in bulk 
and low-dimensional charge-ordered systems, especially in relation to photoinduced phenomena\cite{Yusupov2010,Rohwer2011,Hellmann2012,\mycomment{Monney2016,}Koshihara1999,Schmitt2008,Ichikawa2011,Schmitt2011,Petersen2011,Lee2012,Mertelj2013,Leuenberger2015,Haupt2016,Chaix2017,Laulhe2017,Zong2018,Zong2018a,Kogar2019\mycomment{,Gerasimenko2019}}, we study the photoexcited dynamics of an epitome, the one-dimensional ICDW\cite{Barja2016,\mycomment{Manley2018,Lee2019a,Yang2020,}Monceau2012,Gruener2018}.  
The spatiotemporally varying complex order parameter and the transport and spectroscopic implications reveal exceptionally rich photoinduced nonequilibrium dynamics -- 
the perturbative excitation and propagation of collective amplitude and phase modes and the topological phase transitions due to highly nontrivial phase-winding solitons even for ICDW. The latter emerges as a novel photoinduced topological phenomenon unique to the nonequilibrium and nonlinear system evolution.

\section{Model system with broken continuous symmetry}
\begin{figure}[hbt]
\includegraphics[width=8.6cm]{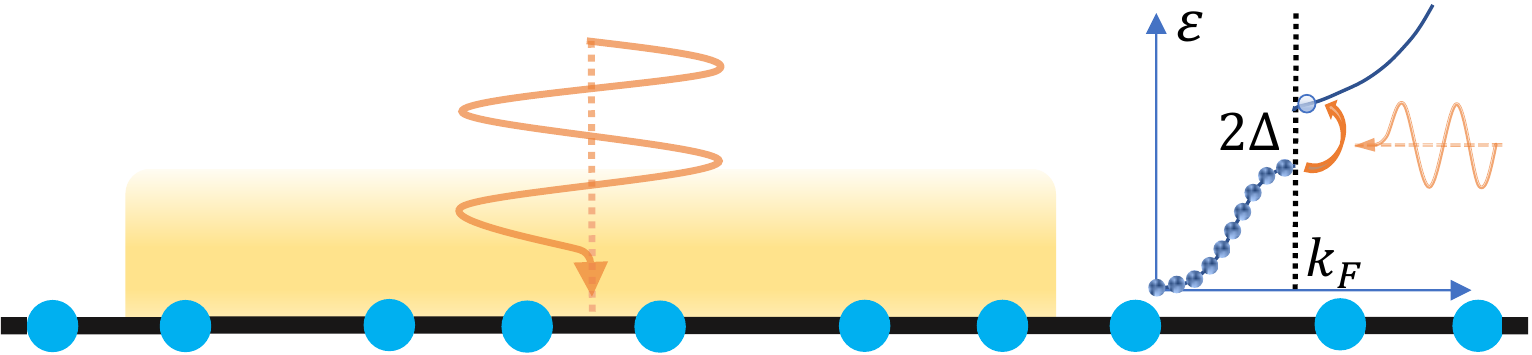}
\caption{Optical irradiation locally onto the yellow region of a chain with ICDW order close to one-third-filling will induce ultrafast nonequilibrium and nonlinear dynamics. Inset: we mainly focus on the consequence of photoinduced electron-hole excitations across the spectral gap $2\Delta$.}\label{Fig:light}
\end{figure}
Fig.~\ref{Fig:light} schematically illustrates our basic setting: an ICDW chain 
is formed as the equilibrium ground state from the electron-phonon interaction, which originally has $N$ ions uniformly spaced at lattice constant $a$. To be specific, we study the Peierls-type electron-phonon coupling Hamiltonian\cite{Barisic1970,Heeger1988,Monceau2012}
\begin{equation}\label{eq:model0_main}
    \mathcal{H} = \mathcal{H}_\mathrm{el}(\vec u) + \mathcal{H}_\mathrm{phn}(\vec u).
\end{equation}
The electronic part $\mathcal{H}_\mathrm{el}(\vec u) = \sum_{i,\sigma}^N  (-t_i c_{i,\sigma}^\dag c_{i+1,\sigma} \mycomment{+ t c_{i,\sigma}^\dag c_{i,\sigma}} + \mathrm{h.c.})$ includes spin-$\sigma$ degeneracy and electron-phonon coupling $\alpha$ through the ion-dependent hopping $t_i=t[1-\alpha(u_{i+1}-u_i)]$; 
the kinetic and potential energies of the phonon displacement field $\vec{u}$ give $\mathcal{H}_\mathrm{phn}(\vec u) = \sum_{i} \frac{M}{2} \dot u_i^2 +  \frac{\kappa}{2}(u_i-u_{i+1})^2$ with the ion mass $M$ and elastic stiffness $\kappa$. Coprime filling fraction representatively as $\frac{p}{q}=\frac{67}{199}\approx\frac{1}{3}$ with a large denominator is used for a long chain $N=10q$, 
approaching the incommensurate limit. 
In equilibrium, the phonons condense at the Kohn anomaly momentum $Q=2k_F\mycomment{=\frac{p}{q}\frac{2\pi}{a}}$ and form a CDW with ionic displacement $u^{(0)}(x)=U\sin{(Qx+\phi_0)}$.
Note that the phenomena of our interest are generic to any ICDW systems, because it is the incommensurability, rather than any specific nonuniversal filling, that is essential to the physical mechanism. We provide extensive examples of other filling fractions in Supplemental Material (SM)\cite{SM}.
\begin{figure*}[hbt]
\includegraphics[width=11cm]{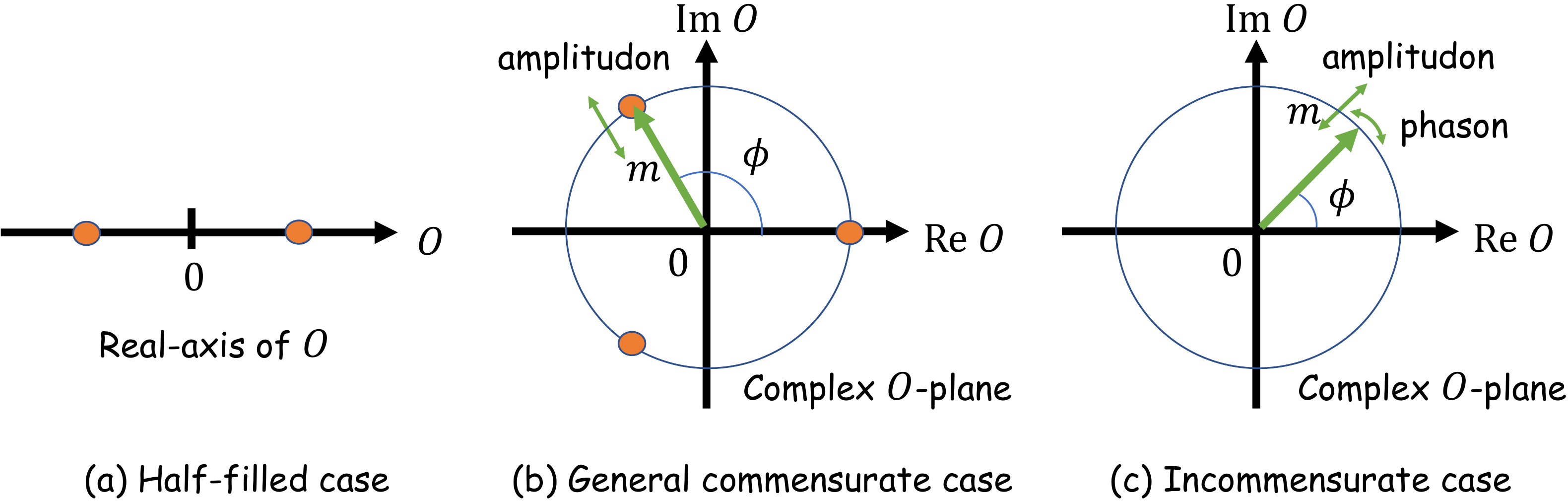}
\caption{Conceptual illustration in the complex order parameter $O$-plane of the CDW symmetry breaking profiles and collective degrees of freedom. (a) Half-filled case leads to a real order parameter of only the amplitude that can take two values denoted by the orange dots. (b) In the general commensurate case, illustrated as a $1/3$-filled case for presentational brevity, $O$ takes three discrete possible values in the complex plane. Only amplitude can fluctuate as a collective excitation. (c) For an incommensurate case, $O$ can continuously take any value on the circle of a given amplitude. Both amplitude and phase can fluctuate as collective excitations.}\label{Fig:SSB}
\end{figure*}

This incommensurate system with broken translational 
symmetry and the associated complex and continuous order parameter $O$ is fundamentally different from any discrete order like in polyacetylene\cite{\mycomment{Su1979,Su1980,}Heeger1988}.
In Fig.~\ref{Fig:SSB}, we schematically describe the significant distinction of ICDW from any half-filled or general commensurate situation, which follows from the spontaneous $\mathrm{U}(1)$ symmetry breaking. Note that only with the incommensurability, which is the key ingredient in our study, the freely varying phase degree of freedom becomes possible. And it, cooperatively together with the amplitude, plays the key role to generate the novel nonequilibrium topological phase.

\section{Nonequilibrium time evolution}
Experimentally an intense light pulse is 
exerted to a relatively local region at time $T<0$ and drives the system via light-matter interaction; notwithstanding an incomplete knowledge of this highly nonlinear and nonequilibrium optical disturbance, we directly focus on the most probable and common photoexcitation -- an electron-hole pair is created at time $T=0$ via absorbing the photon across the spectrum with a CDW gap $2\Delta$\cite{Wall2010,Matsuzaki2015}. 
The phonon dynamical period $\mycomment{t_\mathrm{ad}}\sim T_0= 2\pi/\omega_0$ within an excited energy surface is typically 
more than one orders of magnitude shorter than the interband spontaneous decay ranging from several hundred femtoseconds to picoseconds\cite{Mele1982a\mycomment{,Mele1982},Wall2010,Hellmann2012,Zhang2019b}.
Encoding the pumping effect, we thus take this $T=0$ state as the starting point as well as a given nonequilibrium occupation during the system's 
subsequent evolution 
before the possible (non)radiative recombination decay; 
alternatively, a possibly persistent 
photoirradiation can maintain the excited state beyond free decay time\cite{Su1980}. 
Justified by the hierarchy and separation of two distinct time scales -- electronic and ionic (phonon), such an adiabatic approximation and associated instantaneous electronic states are physically the most relevant since we are interested in the characteristic dynamics or time scale directly related to the phonons’
nonequilibrium behavior. In particular, electrons will in general rapidly relax and adjust to the instantaneous
phonon coordinates at an orders-of-magnitude faster time scale due to the mass difference; as the typical phonon frequency $\mycomment{\hbar}\omega_0=\sqrt{\frac{\kappa}{M}}\ll2\Delta$, off-diagonal electronic density-matrix elements $\rho_{mn}$ oscillate much faster than the lattice and hence negligible electron-phonon energy transfer can interfere the phonon dynamics. 
Indeed, in the following the collective excitations and the
topological transition with soliton formation are inherently in terms of the
phonon coordinate dynamics, upon which we define and study the complex order parameter.

In Appendix~\ref{App:method_evolution}, we outline the simulation of this time-dependent system. Due to the scaling property, the nonequilibrium system possesses two primary parameters: the coherence length $\xi=\frac{\mycomment{\hbar} v_F}{\pi\Delta}$ specifies the electron-phonon system and can be varied from the coupling $\alpha$ in Eq.~\eqref{eq:model0_main} and the 
variationally determined CDW strength $U$; the ratio $\chi=\tau/T_0$ between minimal time step $\tau$ and phonon period for a total evolution time $\mathcal{T}=1000\tau$. 
To faithfully simulate the evolution, we set typically $\chi=1/20\pi$, i.e., phonons complete a full oscillation cycle after more than $60$ steps. Thus, as $T_0\sim60\mathrm{fs}$ in realistic systems, we have $\tau\sim1\mathrm{fs}$ and $\mathcal{T}\sim1\mathrm{ps}$ within or comparable to the foregoing electronic decay time. Henceforth, time $T$ will be measured in units of $\tau$. Lastly, pinning the outset of traveling collective modes and solitons, we put an initial tiny perturbation to the ionic displacement field 
at a few points around the chain center, which as well mimics the local photoirradiation to the chain in Fig.~\ref{Fig:light}. 

\begin{figure*}[hbt]
\includegraphics[width=12.5cm]{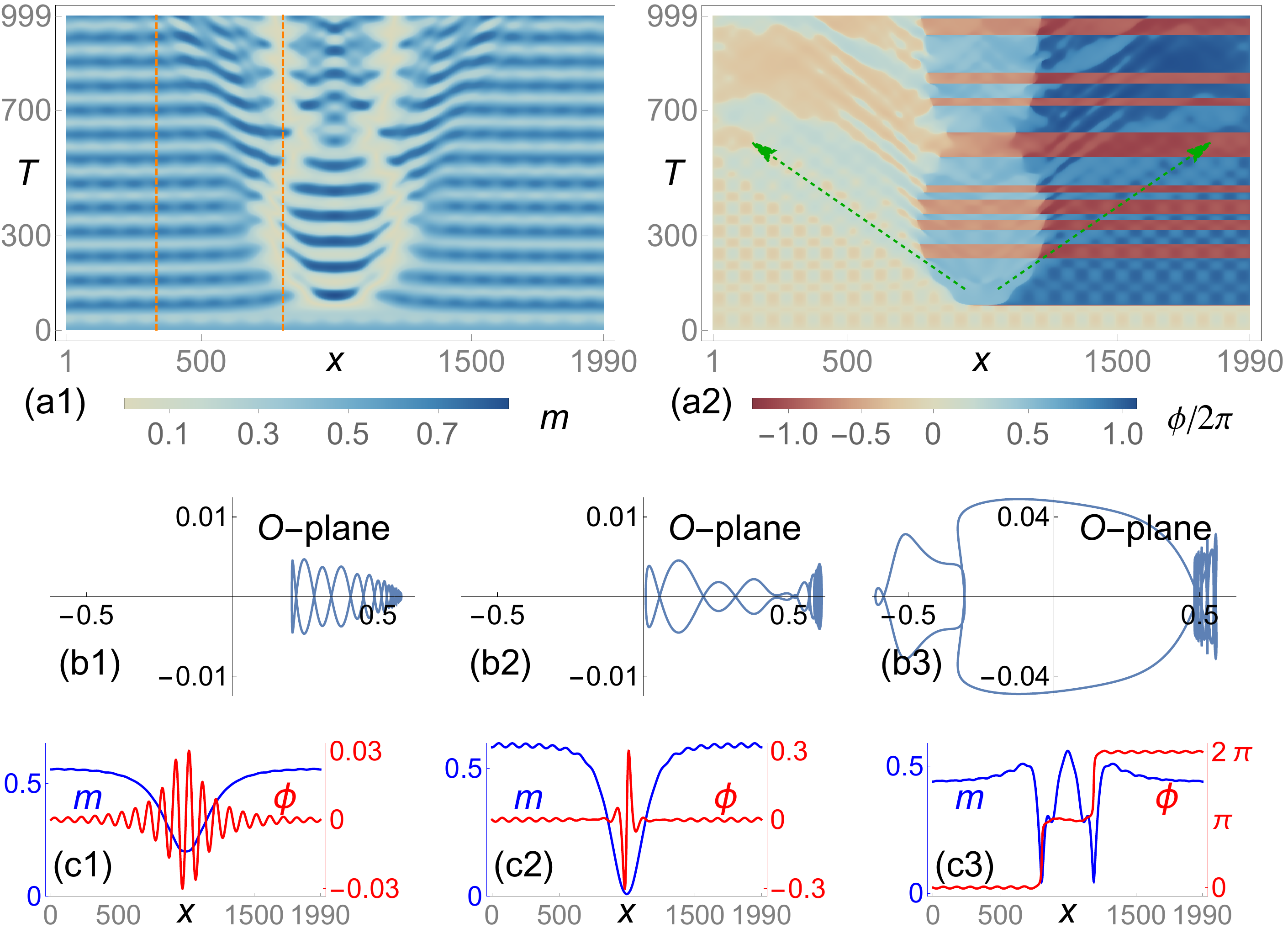}
\caption{(a) Amplitude and phase of the nonequilibrium spatiotemporal ICDW complex order parameter $O(x,T)=m\ee^{\ii \phi}$ at coherence length $\xi\approx 9a$. Horizontal (vertical) axis corresponds to the whole chain (evolution time).
Orange vertical lines at $x=333,801$ in (a1) are used in Fig.~\ref{Fig:spec_cond}(c). 
(b1-3) Complex $O(x)$-trajectories as $x$ traverses the whole chain respectively at $T=71,79,185$ illustrate the topological transition near $T=80$ and the later stabilized solitons. Note the difference in scales between real and imaginary axes.
(c1-3) \mycomment{Snapshots of }$m(x)$- and $\phi(x)$-profiles in one-to-one correspondence to (b1-3).
Parameters are $\hbar=a=t=1,\omega_0=0.006,\alpha=0.55$.}\label{Fig:collective}
\end{figure*}

The most important feature of the nonequilibrium system consists in the instantaneous complex order parameter $O$. As the CDW condensate at momentum $Q$ is stirred by the photoexcitation both perturbatively and nonperturbatively, it in general becomes highly inhomogeneous 
\begin{equation}
    O(x)= m(x) \ee^{\ii\phi(x)} = \int_{-q_0}^{q_0} \dd q \, u_{q+Q} \,\ee^{\ii (q x+\phi_0)} 
\end{equation}
with $u_q$ the momentum-space transform of the instantaneous phonon field $u(x)$ and $q_0=\mycomment{\frac{2\pi}{L}5N_\mathrm{p}=\frac{10\pi}{qa}}10\pi/qa$ a cutoff momentum that stably captures the long-wavelength behavior. 
Shown in Fig.~\ref{Fig:collective}, time-dependent real fields $m(x),\phi(x)$, which are constant $m=U/2,\phi=0$ at equilibrium, become respectively the mass (amplitude) and the phase. While they are decoupled at linear order,  
nonlinear and mutual excitations are naturally possible and incorporated in the simulation, thereby fully encoding both the photoinduced perturbative 
and nonperturbative effects. 
We inspect below these two major aspects specific to ICDW: collective excitation of intrinsic gapless phason and gapped amplitudon 
exhibits rich physical evolution; 
formation of exceptional $\pi$-winding phase solitons, partially captured by low-energy models, realizes new nonequilibrium topological-winding phases. 

\section{Collective excitations}
In Figs.~\ref{Fig:collective}(a), we plot the spatiotemporal profile of amplitude $m$ and phase $\phi$ after the initial photoexcitation. 
Instead of obvious wave propagation, Fig.~\ref{Fig:collective}(a1) shows a robust fast temporal oscillation. This should owe to the gapped amplitudon that gives a vanishing group velocity $\dd\omega_\mathrm{amp}/\dd k|_{k=0}$ and determines the observed standing-wave oscillation frequency $\omega_\mathrm{amp}(k=0)$.
The spacetime V-shape indicated by the green arrows in Fig.~\ref{Fig:collective}(a2) exactly depicts a phason 'light cone' emanating from the chain center where solitons germinate and one can read the gapless phason velocity $v_\mathrm{phs}$ from the envelope shape. The phason wavefront initially splits the spatial pattern and then pushes the phase from the center towards both edges as time elapses.
Note that in the polyacetylene case 
only the amplitude mode exists and the continuous phason degree of freedom associated with the $\mathrm{U}(1)$ order parameter is completely absent. The rich collective excitation behavior in Fig.~\ref{Fig:collective} is only made possible by the incommensurate nature. 

From the mean-field and random-phase-approximation calculation of the collective excitations for ICDW\cite{Lee1974\mycomment{,Gruener2018}}, we obtain the phason velocity $v_\mathrm{phs}=v_F\gamma^{1/2}\omega_Q/2\Delta$ and the amplitude frequency $\omega_\mathrm{amp}(0)=\gamma^{1/2}\omega_Q$ with the dimensionless electron-phonon coupling $\gamma = (\ln{2\pi\xi\varepsilon_F/\mycomment{\hbar}v_F})^{-1}$ and the uncoupled acoustic phonon frequency $\omega_k=2\omega_0|\sin{\frac{1}{2}ka}|$ evaluated at $k=Q$. 
From another viewpoint of practically making a CDW approach incommensurability with large filling denominators, 
the phason and amplitudon originate from the rapidly softened lowest two optical phonons in the Kohn anomaly on the insulating side of the CDW transition\cite{Rice1973,Schulz1978}, which similarly suggests constant $v_\mathrm{phs}/\omega_0a$ and $\omega_\mathrm{amp}(0)/\omega_0$ for a fixed coherence length $\xi$ as the ICDW expression implies. We confirm this general expectation by varying elastic stiffness in our simulation. 
Also note that the observed $\omega_\mathrm{amp}\sim\omega_0$ for our incommensurate system close to $\frac{1}{3}$-filling can be understood via the simpler commensurate $\frac{1}{3}$-filling case, where the amplitudon, originally as the second longitudinal optical phonon branch of frequency $\sqrt{3}\omega_0$ at $q=0$ for a triatomic chain, is lowered around $\omega_0$\cite{Kesavasamy1978,Schulz1978}.
On the other hand, we can further tune the condensate at different coherence lengths, e.g., $\xi_{1,2}/a \approx 3,9$, for which the theory predicts $v_\mathrm{phs2}/v_\mathrm{phs1} \mycomment{= \sqrt{\gamma_2\Delta_1^2/\gamma_1\Delta_2^2}= \sqrt{\frac{\Delta_1^2\ln{\frac{2\varepsilon_F}{\Delta_1}}}{\Delta_2^2\ln{\frac{2\varepsilon_F}{\Delta_2}}}} }\approx 2.5$ and $\omega_\mathrm{amp2}(0)/\omega_\mathrm{amp1}(0)  \mycomment{= \sqrt{\gamma_2/\gamma_1}= \sqrt{\frac{\ln{\frac{2\varepsilon_F}{\Delta_1}}}{\ln{\frac{2\varepsilon_F}{\Delta_2}}}} }\approx 0.86$. As this mean-field prediction can underestimate fluctuations and is valid as to order of magnitude in one dimension\mycomment{\cite{Allender1974}}, it is in reasonable agreement with $v_\mathrm{phs2}/v_\mathrm{phs1} \approx 2.1$ and $\omega_\mathrm{amp2}(0)/\omega_\mathrm{amp1}(0) \approx 0.87$ measured directly from our simulation (see SM\cite{SM}).

\section{Topological phase transition and soliton}
Apart from the collective excitations as a background, there lie more drastic nonperturbative features at $T\gtrsim80\mycomment{\tau}$: the topological soliton-pair formation and its subsequent movement are consistently visible in Figs.~\ref{Fig:collective}(a1,a2) respectively as considerable amplitude $m$-reduction and rapid phase $\phi$-jump. After the initial period when $m(x)$ softens till zero at the chain center, the soliton pair is created and accompanied by a topological transition of the total phase winding along the chain from zero to $\pm2\pi$. 
Plotted in Figs.~\ref{Fig:collective}(b,c), in the complex order parameter $O=m\ee^{\ii\phi}$-plane, this transition means that a photoinduced initially small closed $O(x)$-trajectory deforms to eventually encircle the origin. See also SM\cite{SM} for extended examples, supporting the topological robustness of this origin-winding transition and the insignificance of the specific nonuniversal trajectory shape.
We will later revisit the process with spectral and transport probes, where the associated solitonic states formed inside the CDW gap become clear. After the initial topological transition to a nontrivial phase, 
the horizontal stripes in Fig.~\ref{Fig:collective}(a2) indicate the general possibility of occasional topological switchings between different-winding topological phases. Note that because of the nonintegrable nature of the system, the particular topological-switching pattern and long-time soliton 
behavior are nonuniversal and dependent on the model \mycomment{and simulation }details. 
In reality, solitons with opposite charge can possibly attract each other and eventually get damped or annihilated; here in the absence of an explicit decay channel, they persist and can contingently collide.

The exact topological winding quantization mostly consists of 
two separate $\pi$-solitons [Figs.~\ref{Fig:collective}(b3,c3)], where amplitude $m$ is reduced but remains finite at the soliton position, distinct from the amplitude/mass sign-change solitons or alike edge states. Instead, the novel nonequilibrium topological soliton herein is stabilized by taking advantage of the phase freedom and its winding quantization. Only right at a topological phase transition, including the initial one  [Figs.~\ref{Fig:collective}(b2,c2)] and later topological switchings between $\pm 2\pi$ [see Figs.~S11,S12 in SM]
can the solitons become singularly sharp and $m\rightarrow0$. We also show in SM higher-winding transitions as a general feature of more than one electron-hole photoexcitations. The emergence of (multiples of) $\pi$-solitons needs to be understood as a nontrvial feature of the ICDW, because it naturally holds the continuous phase freedom aforementioned in contrast to polyacetylene's staggered order or any commensurate case. Instead, the incommensurability eventually suppresses all the possible commensurate multiples of $2p\pi/q$-windings\cite{Rice1976,Su1981} but singles out $\pi$-phase-kink solitons to compose the total winding quantization for the nonequilibrium ICDW, which is fundamentally different from polyacetylene's amplitude soliton.
\begin{figure}[hbt]
\includegraphics[width=8.6cm]{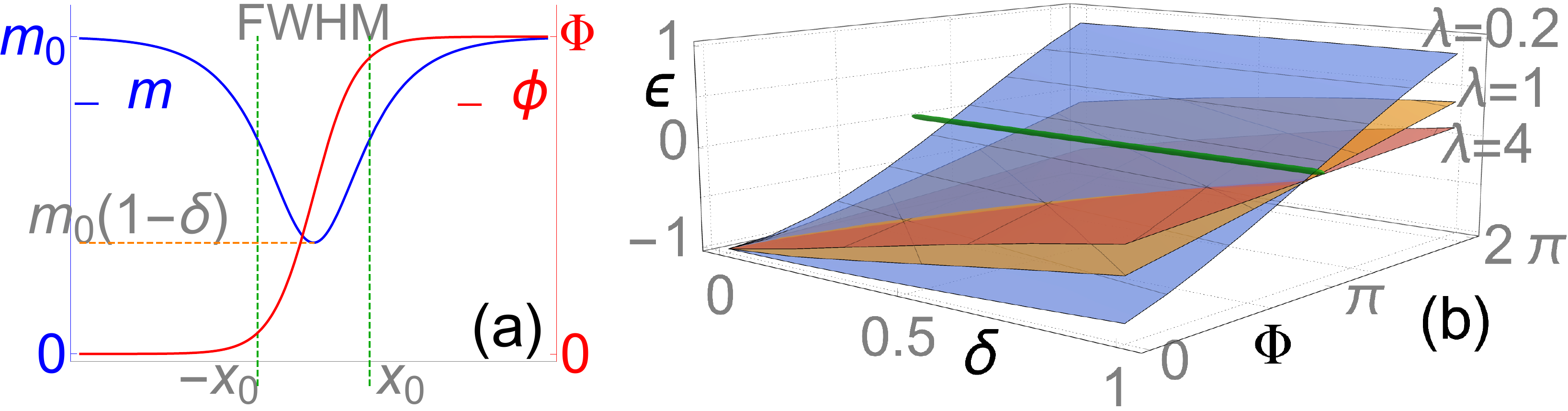}
\caption{(a) Soliton profile centered at $x=0$ of amplitude $m$ and phase $\phi$ with relative weakening $\delta$, phase-winding $\Phi$, and full-width-half-maximum (FWHM) $2x_0=2\lambda\sech^{-1}\!{\frac{1}{2}}$. (b) Soliton energy $\epsilon(\lambda,\delta,\Phi)$ solved from the continuum model. Green line: Jackiw-Rebbi zero mode.}\label{Fig:soliton}
\end{figure}

We can derive from Eq.~\eqref{eq:model0_main} a low-energy continuum model to partially capture the in-gap soliton state (see Appendix~\ref{App:continuum})
\begin{equation}\label{eq:solitonequ_main}
    h = v_F (-\ii \partial) s_z + \beta [O(x) s_+ + O^*(x)s_-],
\end{equation}
where $\beta=2\alpha v_F/a$. The pseudospin 
$s_z,2s_\pm=s_x\pm\ii s_y$ combines the electrons around two Fermi points and real spin degeneracy is assumed. 
Resembling the realistic profiles like Fig.~\ref{Fig:collective}(c), the complex order parameter $O(x)$ enters for a soliton as $m(x)=m_0 (1-\delta\sech{\frac{x}{\lambda}}),\phi(x)=\frac{\Phi}{2} \tanh(\frac{x}{\lambda}+1)$, where $\lambda,\delta,\Phi$ control respectively the soliton width, amplitude weakening and phase winding
in Fig.~\ref{Fig:soliton}(a). Scaling properties of $h$ makes it suffice to set $v_F=m_0=\beta=1$ and have $x,\lambda$ measured in units of  
$\lambda^*=v_F/m_0\beta=a/2\alpha m_0\approx60a$ for Fig.~\ref{Fig:collective}.
This soliton characteristic scale $\lambda^*$, in comparison to the condensation momentum $Q$, signifies the length scale separation between the slowly varying envelope and the rapidly oscillating CDW pattern and hence justifies the continuum theory and the cutoff-independent long-wavelength description of the inhomogeneous $O$.
Although this low-energy $h$ suffers spectral pollution\cite{Lewin2009\mycomment{,Almanasreh2012}}, we are able to solve its bound-state energies $\epsilon\in(-m_0,m_0)$ by the compound-matrix method relying on a topological invariant of the differential system\cite{\mycomment{Jones1990,Allen2002,}Pearce2010}.

In Fig.~\ref{Fig:soliton}(b), we plot energy $\epsilon$ lowest in absolute value and find it monotonically increasing with $\delta,\Phi$, which is the low-energy excitation around Fermi points and enables energy gain from soliton formation. 
As $\lambda$ shrinks, the variation against $\delta(\Phi)$ becomes flatter (steeper) and the $\epsilon=0$ solutions move more aligned to $\Phi=\pi$. 
This is asymptotically consistent with the Jackiw-Rebbi zero mode, because solutions become insusceptible to $\delta$ when $\lambda\rightarrow 0$. For typical smooth solitons $\lambda\lesssim 1$ [Fig.~\ref{Fig:collective}(c3)], though finite $\delta$ costs elastic energy due to the amplitude deformation, it 
helps to form near-zero-energy solitons: phase-winding and amplitude-reduction together assist in stabilizing the photoexcitation. 
Single $2\pi$-solitons can have nearly zero energy only when $\lambda$ is long enough, which indeed is the rare case at nonequilibrium because of the higher elastic-energy penalty accumulated along the wider distortion. 
The total $2\pi$-winding is more commonly achieved by two separate sharper kinks of $\pi$-winding [Fig.~\ref{Fig:collective}(c3)], because, except the Jackiw-Rebbi resemblance for sharper solitons, the most energy gain from bound-state formation, $|m_0-\epsilon(2\pi-\Phi)|+|-m_0-\epsilon(\Phi)|$, is 
maximized around $\Phi=\pi$ following the monotonicity of $\epsilon$. 
Therefore, the robust nonequilibrium topological soliton in ICDW, for energetic and topological reasons, 
nontrivially embodies a $\pi$-winding spinon soliton with neutral charge accumulation as per the single occupation of electron-hole pair.

\section{Transport and spectral probes}
\begin{figure*}[hbt]
\includegraphics[width=17.8cm]{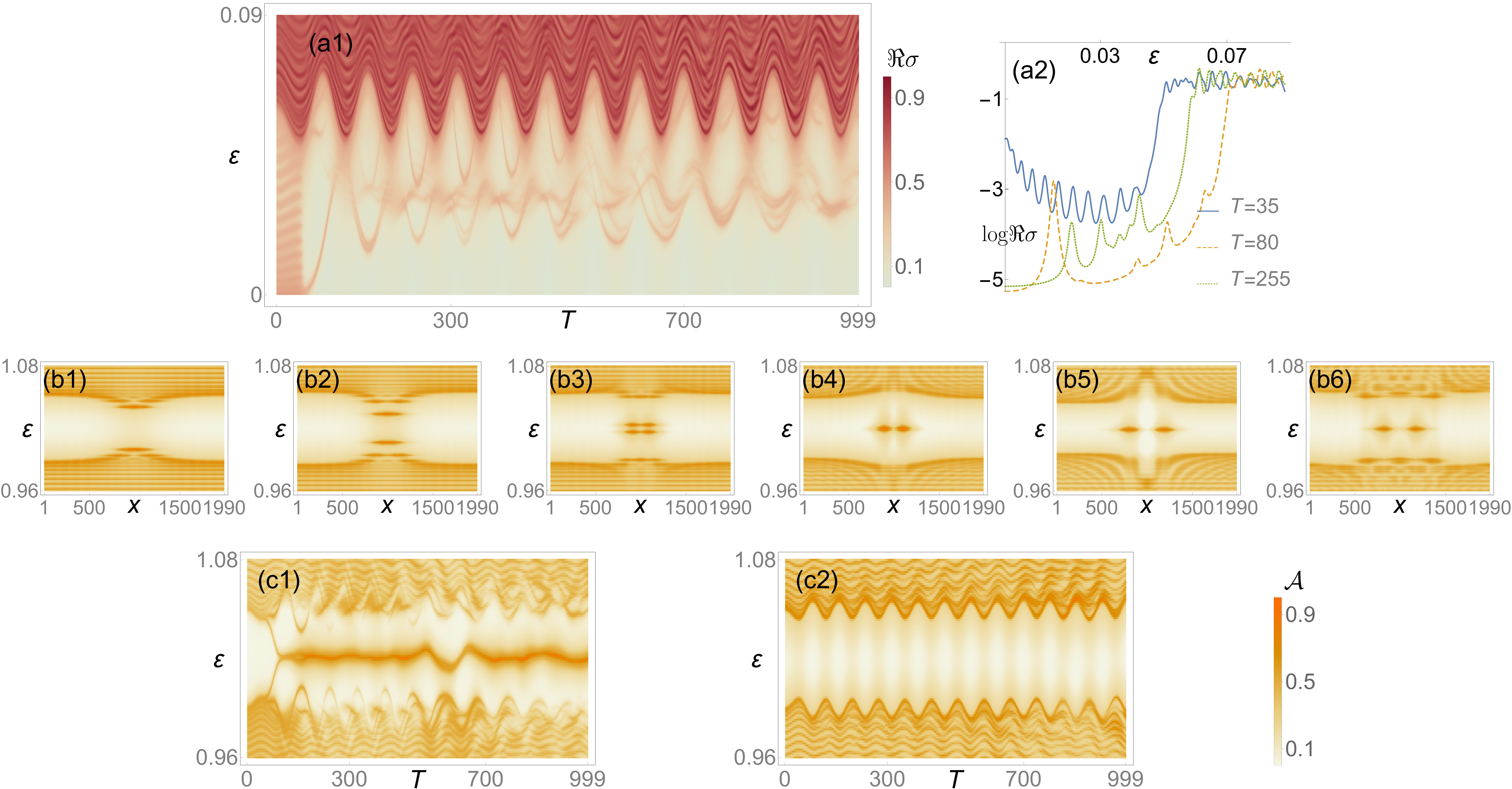}
\caption{(a1) Real part of the time-dependent optical conductivity $\Re\sigma(\varepsilon,T)$. (a2) Snapshots of $\log\Re\sigma(\varepsilon,T)$ at $T=35,90,255$. (b1-6) Space-dependent spectral function $\mathcal{A}(\varepsilon,x)$ respectively at times $T=65,75,90,100,200,700$. (c1-2) Time-dependent spectral function $\mathcal{A}(\varepsilon,T)$ at sites $x=801,333$ corresponding to the orange lines in Fig.~\ref{Fig:collective}(a1). All quantities, calculated from the case of Fig.~\ref{Fig:collective} with $\eta=10^{-3}t$ comparable to the level spacing, are scaled with respect to their maxima. See SM for extended early-time data.}\label{Fig:spec_cond}
\end{figure*}
Time-resolved optical conductivity $\sigma$ and spectral function $\mathcal{A}$ 
constitute the comprehensive experimental access towards 
dynamical properties of this nonequilibrium system. 
Conductivity can be extracted from transport and terahertz pump-probe optical reflection or transmission measurement\cite{\mycomment{Fox2010,}Kuzmenko2005,Onda2008,Dienst2011,Hu2014,Singla2015,Mitrano2016,Nicoletti2016}. Spectral weight, while directly corresponding to local or integrated density of states and hence STM tunneling I-V spectra that can resolve in-gap solitons\cite{Cocker2013,Eisele2014,Yoshida2013,Yoshida2014,Cocker2016}, is also a target of spatially-resolved nano-ARPES using synchrotron lights\cite{Giannetti2016,Lv2019,Cattelan2018}.
We derive in Appendix~\ref{App:Kubo}
\begin{equation}
    \sigma(\varepsilon,T)
    = -\frac{\ii e^2}{\mycomment{\hbar} L} \sum_{mn} \frac{(f_n-f_m)(\varepsilon_n-\varepsilon_m) |x_{m n}|^2}{\mycomment{\hbar}(\varepsilon +\ii\eta)+\varepsilon_n-\varepsilon_m}
\end{equation}
with the instantaneous electronic spectrum and states $\{\varepsilon_i,\ket{i}\}$ 
and the nonequilibrium occupation $f_i$ 
directly from the simulation and the position operator $x_{m n}=\langle m \lvert x \rvert n\rangle$ 
for the aperiodicity after photoexcitation. 
Alongside, we calculate 
the time- and space-dependent 
$\mathcal{A}(\varepsilon,x,T)=-\frac{1}{\pi}\Im\mathcal{G}^\mathrm{r}(\varepsilon,x,x,T)$ from the retarded electronic Green's function $\mathcal{G}^\mathrm{r}$.

We observe in Fig.~\ref{Fig:spec_cond}(a1) that below the dark-red conductivity continuum due to across-gap excitations, there lies the region ($\varepsilon\lesssim 2\Delta$) exhibiting reddish nonzero signals only present due to the photoexcitation and in-gap topological states. As the evolution starts, the excited electron and hole at the conduction/valence band edges, absorb photon\mycomment{aided by the driving electric field} and transition to many nearby empty or filled states [earlier than but similar to Fig.~\ref{Fig:spec_cond}(b1)]; hence the dense reddish signal starting from $\varepsilon=0$ in Figs.~\ref{Fig:spec_cond}(a1,a2) within the conductivity gap 
for a short time $T\lesssim50$, prior to the soliton maturation or topological transition $T\approx80$ in Fig.~\ref{Fig:collective}. Within this period, around the chain center the originally \mycomment{spatially }uniform band edges deform towards the gap and prepare the soliton germination [Figs.~\ref{Fig:spec_cond}(b1-2)]. Later, the electron-hole states gradually split off the band continuum and move deep inside the gap, developing the spatially separated double $\pi$-winding structure and finally transforming into the stabler solitons \mycomment{energetically }around the gap middle [Figs.~\ref{Fig:spec_cond}(b2-4)]; conductivity now mainly originates from the partially filled soliton states and those unstable states around band edges -- for instance, the upward thin-line ($50<T<120$) emanating from the $\varepsilon=0$ bottom\mycomment{ of the aforementioned dense in-gap signal} in Fig.~\ref{Fig:spec_cond}(a1) is primarily attributed to the increasing energy separation between solitons and band edges.

During the later time evolution, while \mycomment{relatively }stabilized solitons could still fluctuate\mycomment{ at late times}, band-edge states constantly bear variation in energy and spectral weight\mycomment{ both at initial and late times} [Figs.~\ref{Fig:spec_cond}(b2-6)], which can even slightly detach transient states temporarily into the gap, complicating the conductivity signals.
However, major in-gap signals roughly around the conductivity gap midpoint $\varepsilon\approx \Delta$ in Fig.~\ref{Fig:spec_cond}(a1) clearly reflect the contribution from the solitons \mycomment{more stabilized }around the middle of the CDW gap and the apparently oscillating band edges seen through Figs.~\ref{Fig:spec_cond}(b). This regular gap oscillation becomes conspicuous in the time-dependent spectral function at a site away from the soliton trajectory, otherwise obscured by the soliton contribution at a site traversed by the soliton [Figs.~\ref{Fig:spec_cond}(c)].
Correspondingly, it is clearly (vaguely) picked up in Fig.~\ref{Fig:spec_cond}(a1) by the dark-red continuum (the reddish in-gap signals). This temporal oscillation in both $\sigma$ and $\mathcal{A}$ is the direct consequence of the photoexcited amplitudon and perfectly matches the oscillation in Fig.~\ref{Fig:collective}(a1).
Furthermore, Fig.~\ref{Fig:spec_cond}(c2) prevails in excluding the soliton spectral weight whereas Fig.~\ref{Fig:spec_cond}(c1) recapitulates the soliton energetic evolution that first leaves the 
band edges and eventually stabilizes inside the gap. The temporally light-heavy-light-heavy in-gap weight in Fig.~\ref{Fig:spec_cond}(c1) exactly records the 8-shaped soliton trajectory crossing the orange line in Fig.~\ref{Fig:collective}(a1). 

\section{Discussion}
Here, we comment on the physical validity of the approach we adopt in relation to the pump-probe-like spectroscopy, admittedly one of the relevant and practical experiment for realizing the proposal of nonequilibrium phenomenon. 
We make use of the adiabatic approximation, which has been widely used and is physically justified for the ion/phonon nonequilibrium dynamics of our interest. In this regard, at the time scale or resolution for experimentally monitoring the leading nonequilibrium dynamics of interest, the system has very rapid electronic transition: the electronic states effectively adjust to the phonon coordinates instantaneously, during which phonons mainly remain intact. Explicitly incorporating the electromagnetic pump-probe field does not alter the physical content of the optical conductivity and spectral function at this time scale of practical and theoretical interest. This justifies that instantaneous eigenstates suffice to capture the experimentally relevant time-dependent signals. 

As for the initial optical disturbance or pumping that gives rise to the electron-hole excitation, the optical absorption especially happens at a much higher energy scale with the lower bound given by the large CDW gap, where the transition time is correspondingly much shorter than the interested phonon dynamics. Therefore, again at the time scale relevant to the nonequilibrium dynamics of our interest, encoding the optical pumping effect by the electron-hole excitation is sufficient and physically most relevant. This is also in the same spirit as the established quenched dynamics in, e.g., superconducting systems\cite{Barankov2006,*Yuzbashyan2006,Shimano2020}. The emphasis is directly put on the instant change of the pairing instead of an all-encompassing time-dependent electromagnetic pump-probe description, although experimentally it is realized in pump-probe spectroscopy. Such theories capture the experimentally most relevant collective excitations, such as the twice-gap oscillations in both conventional and unconventional superconductors. We hence 
expect our theory to play a similar role in capturing the important nonequilibrium phenomena in a straightforward manner and to stimulate further interest and study.

On the other hand, the explicit pump-probe approach would bring about much theoretical and technical difficulty. In the present case, as incommensurability is of central significance to the phason phenomena, we are confronted by a nonequilibrium correlated CDW system of three main difficulties: the lack of momentum as a good quantum number, an inevitably large enough inhomogeneous system with many degrees of freedom of the full phonon and electron spectra, and the essential nonperturbative effect. This is in stark contrast to the usual incorporation of a pump-probe process: it is more feasible to simpler and smaller systems such as a single band (and typically single-frequency phonon) with a quench or uniform pump-probe field and often limited to perturbation or other approximation, which cannot capture the novel real-space nonperturbative topological phenomena in our system\cite{Kemper2015,Xu2019}. This type of explicit pump-probe treatment has a different focus and advantage. 

Instead, we in this study succeed in providing the way to overcome these difficulties and tackle interesting systems hitherto inaccessible from other methods. And in the spirit aforementioned, we extract the most relevant nonequilibrium phenomena by directly targeting at our large system right after some pumping transition that absorbs the photon and creates the particle-hole excitation, which is the most ordinary effect of the optical disturbance in this system gapped by the CDW order. 
In summary, we combine ICDW as a prototypical quantum phase bearing broken continuous symmetry with the facet of optically induced nonequilibrium and nonlinearity. Real-time ultrafast evolution highlights the spatiotemporally varying complex order parameter and the experimental implication, revealing unique features from the continuous amplitude and especially the phase variation that leads to rich collective dynamics and highly nontrivial solitonic topological phases and transitions unique to nonequilibrium.

\appendix

\section{Simulation of time evolution}\label{App:method_evolution}
Instead of the 
involved full coupled quantum dynamics 
in a large phase space, we treat the system semiclassically: 
electrons effectively integrated out at each instant generate adiabatic forces dominating short-time dynamics, whereby we evolve the ionic configuration. 
The instantaneous electronic 
spectrum and eigenstates $\{\varepsilon_i,\ket{i}\}$ solved from $\mathcal{H}_\mathrm{el}(\vec u)$ lead to the electronic energy $E_\mathrm{el}[\vec{u}]=\sum_{i} f_{i} \varepsilon_i$ where the nonequilibrium occupation $f_i$ incorporates spin degeneracy and keeps the electron-hole pair excited at the band edge. 
Variation of the total energy $E_\mathrm{tot}[\vec{u}]$, combining $E_\mathrm{el}$ and the potential part of $\mathcal{H}_\mathrm{phn}(\vec u)$, determines the conservative force $F_i[\vec{u}] = -\frac{\delta E_\mathrm{tot}}{\delta u_i}$ felt by the $i$th ion. The phonon dynamics 
can thus be performed 
using the velocity Verlet method, 
preserving the symplectic structure and is energy-drift-free.

Based on Eq.~\eqref{eq:model0_main}, the electronic part $\mathcal{H}_\mathrm{el}$ can be solved at any given displacement configuration $\vec{u}$, leading to energy levels $\varepsilon_{i=1,\cdots,N}$. The electronic energy then reads $E_\mathrm{el}[u]=\sum_{i} f_{i} \varepsilon_i$. In the ground state, the occupation number $f_{i}=2$ for the two spins when $i=1,\cdots,N_\mathrm{e}/2$ and $f_i=0$ otherwise. If the light excites an electron from an fully occupied level $j$ to an unoccupied level $k$, i.e., an electron-hole pair, it differs from the ground state in $f_j=f_k=1$. In general, we mostly focus on the scenario of exciting an electron-hole pair right across the CDW gap, i.e., $j=\frac{p}{q}N=N_\mathrm{e}/2,k=j+1$. On the other hand, the phonon potential energy $V_\mathrm{phn}[\vec{u}]=\sum_i \frac{\kappa}{2}(u_i-u_{i+1})^2$.
We thus define a total 'potential' energy $E_\mathrm{tot}[\vec{u}]=E_\mathrm{el}[\vec{u}] + V_\mathrm{phn}[\vec{u}]$ that directly determines the conservative force $F_i[\vec{u}] = -\frac{\delta E_\mathrm{tot}}{\delta u_i}$ felt by the ion $i$. Any ground state must be determined for a fixed $\alpha$ by the condition $\frac{\delta E_\mathrm{tot}}{\delta U}=0$ that energetically minimizes the system for the given equilibrium CDW pattern $u^{(0)}_n= U \sin (Qna+\phi_0)$, where we set the spontaneously broken phase $\phi_0=0$ without loss of generality. When other control parameters are fixed, $\alpha$ will determine $U,\Delta,\xi$. As concrete values of $\alpha,U,\Delta$ are not of interest, we primarily tune $\alpha$ to vary at will $\xi$ in units of $a$, because it fully represents the CDW state's feature. 
Without loss of generality, we impose the periodic boundary condition in the calculation. We also confirm all essential results with the open boundary condition.

Regarding the nonequilibrium time evolution, we integrate the phonon dynamics $M \ddot{\vec{u}} = \vec F[\vec{u}]$ in time steps of $\tau$ using the velocity Verlet method in molecular dynamics\mycomment{\cite{Hairer2006}} that updates both $\vec{u}$ and its velocity consecutively
\begin{equation}\label{eq:Verlet}
\begin{split}
    \vec{u}^{(i+1)}=\vec{u}^{(i)} + \vec{v}^{(i)}\tau + \frac{\vec{F}^{(i)}}{2M}\tau^2\\
    \vec{v}^{(i+1)}=\vec{v}^{(i)} + \frac{\vec{F}^{(i)}+\vec{F}^{(i+1)}}{2M}\tau
\end{split}
\end{equation}
where the force field $\vec{F}^{(i)}$ at time $T=i\tau$ is solved from variating the instantaneous electron system $\mathcal{H}_\mathrm{el}^{(i)}$ against the ion field $\vec u^{(i)}$. This method 
is of second-order global error.
As mentioned in the main text, the whole time-dependent system possesses two independent dimensionless parameters: $\xi/a$ and $\chi=\tau/T_0$ for a fixed number of time steps. 
To see this, we look at a few scaling properties below. Derived from Eq.~\eqref{eq:model0_main}, when $\xi/a$ is invariant one has
\begin{equation}\label{eq:scale}
\begin{split}
    E_\mathrm{tot}(r,s) = rE_\mathrm{tot}(1,1),\,
    \vec{F}(r,s) = \frac{r}{s}\vec{F}(1,1),
\end{split}
\end{equation}
where $E_\mathrm{tot}(r,s) \equiv E_\mathrm{tot}(rt,\frac{r}{s^2}\kappa,s\vec{u},\frac{\alpha}{s})$ and similarly $\vec{F}(r,s)$ are defined for a system with rescaled parameters. Note that $\Delta$ scales the same as $E_\mathrm{tot}$ does. 
Alternatively, this can be seen as altering the units of energy and length.
On the other hand, when one integrates the phonon dynamics via Eq.~\eqref{eq:Verlet},
\begin{equation}\label{eq:ph_integrate}
\begin{split}
    \vec u^{(n)}&=\vec u^{(0)}+n\vec v^{(0)}\tau + \frac{\tau^2}{2M}[n\vec F^{(0)} + 2\sum_{i=1}^{n-1} (n-i)\vec F^{(i)}] \\
    \vec v^{(n)}&=\vec v^{(0)}+\frac{\tau}{2M}[\vec F^{(0)}+2\sum_{i=1}^{n-1} \vec F^{(i)}+\vec F^{(n)}].
\end{split}
\end{equation}
The force field $\vec{F}[\vec u]$ along the chain is purely a functional of the ion displacement field $\vec{u}$. Therefore, when the ions are initially at rest $\vec v^{(0)}=0$, which is the present case, either $\vec{F}^{(n)}[\vec u]$ or the ion displacement from its initial value $\vec u^{(n)}-\vec u^{(0)}$ (implicitly) depends on the combination $\frac{\tau^2}{M}=\frac{(2\pi\chi)^2}{\kappa}$. We immediately observe from Eqs.~\eqref{eq:scale}\eqref{eq:ph_integrate} that as long as $\chi$ is fixed, i.e., $\frac{\tau^2}{M}\rightarrow\frac{\tau^2}{M}\frac{s^2}{r}$ under the scaling, the evolution of $\vec{u}$ is merely rescaled to be $s\vec{u}$ and hence no essential difference although one uses completely different $t,\kappa,\alpha$.
Besides, given other parameters, the time evolution is invariant when one tune $M$ and $\tau$ such that $\frac{\tau^2}{M}$ is fixed, which also means that $\omega_0$ can be tuned 
to fulfill the adiabatic condition $\mycomment{\hbar}\omega_0\ll2\Delta$. Readily seen, the relevance of a concrete choice of $M,\tau$ enters when we calculate the velocities of collective excitations, which involves the time lapse in units of $\tau$ and the space traversed in units of $a$. One has to fix a realistic $M$ and hence $\tau$ in order to compare these velocities with, say, a realistic Fermi velocity $v_F$. See SM for other simulation details.

\section{Low-energy continuum model}\label{App:continuum}
The complete second-quantization form of Eq.~\eqref{eq:model0_main} is
\begin{equation}\label{eq:model1_main}
\begin{split}
    \mathcal{H} &= \sum_{k,\sigma}\varepsilon_k c_{k\sigma}^\dagger c_{k\sigma}+\sum_q\omega_q b_q^\dagger b_q\\
    &+\sum_{k,q,\sigma}(g_{k,q}c_{k+q,\sigma}^\dagger c_{k\sigma}(b_q+b_{-q}^\dag)+\mathrm{h.c.})
\end{split}
\end{equation}
wherein $c_{k\sigma}=\frac{1}{\sqrt{N }}\sum_n\ee^{-\ii kna}c_{n\sigma}$, $u_n=\sum_q\sqrt{\frac{\hbar}{2N M\omega_q}}(b_q+b_{-q}^\dag)\ee^{\ii qna}$, and the electron-phonon vertex $g_{kq}=t\alpha\sqrt{\frac{\hbar}{2N M\omega_q}}\ee^{\ii ka}(\ee^{\ii qa}-1)$. 
Once the phonons in Eq.~\eqref{eq:model1_main} are integrated out, the electron-phonon coupling will effectively induce attractive four-fermion interactions and hence a BCS-type theory of the CDW pairing. More explicitly, Eq.~\eqref{eq:model1_main} becomes at the mean-field level 
\begin{equation}\label{eq:model2}
    \mathcal{H} = \sum_{k,\sigma}\varepsilon_k c_{k\sigma}^\dagger c_{k\sigma}+\sum_q\omega_q \braket{b_q^\dagger b_q} + \sum_{k,q,\sigma}(f_{k,q} c_{k+q,\sigma}^\dagger c_{k\sigma}+\mathrm{h.c.})
\end{equation}
where we define $f_{k,q} = g_{k,q} \braket{b_q+b_{-q}^\dag}$. Taking into account the phonon condensation at $q=\pm Q$, the linearized low-energy theory around the Fermi level $\varepsilon_F$ is given by setting $k\rightarrow\pm k_F+k,q\rightarrow\pm (Q+q)$ and then summing over $k$ ($q$) that is henceforth small and measured relative to $\pm k_F$ ($\mp Q$)
\begin{equation}\label{eq:model3}
\begin{split}
    \mathfrak{h}  
    \mycomment{&= \sum_{k\rightarrow -k_F+k,q} \varepsilon_k c_{k}^\dagger c_{k} + (f_{k,Q+q} c_{k+Q+q}^\dagger c_{k} + f_{k,-Q-q}^* c_{k}^\dagger c_{k-Q-q})  + \sum_{k\rightarrow k_F+k,q} \varepsilon_k c_{k}^\dagger c_{k} + (f_{k,-Q-q} c_{k-Q-q}^\dagger c_{k} + f_{k,Q+q}^* c_{k}^\dagger c_{k+Q+q}) \\}
    &= \sum_{s=\pm,k\rightarrow sk_F+k} [ \varepsilon_k c_{k}^\dagger c_{k} + \sum_q (f_{k,-s(Q+q)} c_{k-s(Q+q)}^\dagger c_{k} \\
    &+ f_{k,s(Q+q)}^* c_{k}^\dagger c_{k+s(Q+q)}) ] \\
    \mycomment{&= \sum_{k\rightarrow -k_F+k,q} \varepsilon_k c_{k}^\dagger c_{k} + (f_{k,Q+q} + f_{k+Q+q,-Q-q}^*) c_{k+Q+q}^\dagger c_{k} + \sum_{k\rightarrow k_F+k,q} \varepsilon_k c_{k}^\dagger c_{k} + (f_{k,-Q-q} + f_{k-Q-q,Q+q}^*) c_{k-Q-q}^\dagger c_{k} \\}
    &= \sum_{s=\pm,k\rightarrow sk_F+k} [ \varepsilon_k c_{k}^\dagger c_{k} \\
    &+ \sum_q (f_{k,-s(Q+q)} + f_{k-s(Q+q),s(Q+q)}^*) c_{k-s(Q+q)}^\dagger c_{k} ] \\
    &= \sum_{k} [\sum_{s=\pm} \varepsilon_{s,k} c_{s,k}^\dagger c_{s,k} + \sum_{q}(\mathcal{O}_{k,q} c_{+,k_+}^\dagger c_{-,k_-}+\mathrm{h.c.}) ] 
\end{split}
\end{equation}
where we drop the constant term and the spin indices for simplicity, shift the reference energy to $\varepsilon_F$ by $\varepsilon_{\pm,k}=\varepsilon_{\pm k_F+k}-\varepsilon_F=\pm v_F k$, shift $k\rightarrow k+s\frac{q}{2}$ respectively for $k$ in $s=\pm$ branches to univocally define $k_s=k+s\frac{q}{2}$, and $\mathcal{O}_{k,q}=f_{-k_F+k_-,Q+q} + f_{k_F+k_+,-Q-q}^*$. The $k_s$-dependence, $\ee^{\pm\ii k_s a}$, in $\mathcal{O}_{k,q}$ at most generates elements tridiagonal in the real-space lattice representation. The continuum limit near $\pm k_F$ and the long wavelength limit of $q$ ($a\rightarrow 0$) assure of dropping this $k_sa$ and $qa$ dependence and thus we replace $\mathcal{O}_{k,q}$ by a complex mass 
\begin{equation}
\begin{split}
    \mathcal{O}_q&=f_{-k_F,Q+q} + f_{k_F,-Q-q}^* 
    \mycomment{&= 2t\alpha\frac{1}{\sqrt{N}}u_{Q+q}\ee^{-\ii k_Fa}(\ee^{\ii Qa}-1) \\}
    = \ii 4t\alpha\sin{(k_Fa)}\frac{1}{\sqrt{N}}u_{Q+q}.
\end{split}
\end{equation}
For the equilibrium state of a constant mass $\mathcal{O}=\mathcal{O}_{q=0}$ only, from Eq.~\eqref{eq:model3}, we thus have in the real space $h^{(0)} = v_F (-\ii \partial) s_z + \mathcal{O}_0 s_+ + \mathcal{O}_0^*s_-$ with the pseudospin $s_z,s_\pm=\frac{1}{2}(s_x \pm \ii s_y)$ for the two Fermi points. For the more general case with perturbations and soliton formation in nonequilibrium, we obtain from Eq.~\eqref{eq:model3}
\begin{equation}\label{eq:model4}
\begin{split}
    \mathfrak{h}
    &\approx \sum_{k} [\sum_{s=\pm} \varepsilon_{s,k} c_{s,k}^\dagger c_{s,k} + \sum_{q}(\mathcal{O}_{q} c_{+,k_+}^\dagger c_{-,k_-}+\mathrm{h.c.}) ] \\
    &= \int \dd x [\sum_{s=\pm} s v_F (-\ii\partial) c_{s,x}^\dagger c_{s,x} + (\ii\beta O(x) c_{+,x}^\dagger c_{-,x}+\mathrm{h.c.}) ]
\end{split}
\end{equation}
wherein $\beta=4t\alpha\sin{(k_Fa)}$, we have used $\sum_k{\ee^{\ii k(x-x')}}=N\delta_{x,x'}\to\delta(x-x)$ for transitioning to the continuum limit, and we can now define the complex mass or order parameter
\begin{equation}\label{eq:OP}
    O(x)=m(x)\ee^{\ii\phi(x)}=\sum_q{\frac{1}{\sqrt{N}}}u_{Q+q}\ee^{\ii q x}    
\end{equation}
that no longer remains a constant. We use the Fourier transform $u_k=\frac{1}{\sqrt{N }}\sum_x\ee^{-\ii kx}u(x)$. Real quantities $m(x),\phi(x)$ take values $m=U/2,\phi=\phi_0$ at equilibrium. 

Therefore, we finally have the Hamiltonian
\begin{equation}\label{eq:solitonequ}
    h = v_F (-\ii \partial) s_z + (\ii\beta O(x) s_+ + \mathrm{h.c.})
\end{equation}
appllied on a wavefunction $\psi(x)=(\psi_+(x),\psi_-(x))^\mathrm{T}$.
Using the concrete form of $O(x)=m_0\tilde{O}(\frac{x}{\lambda})$ with a characteristic length scale $\lambda$, we first nondimensionalize Eq.~\eqref{eq:solitonequ} 
\begin{equation}
    \tilde{h}  = -\ii \partial_{\tilde{x}} s_z + 2\tilde\alpha \tilde{m}_0(\ii \tilde{O}(\frac{\tilde{x}}{\tilde\lambda}) s_+ + \mathrm{h.c.})
\end{equation}
where all tilde quantities $\tilde{x}=x/a,\tilde{\lambda}=\lambda/a,\tilde{\alpha}=a\alpha,\tilde{m}_0=m_0/a$, $\tilde{h}=h/\epsilon_0$ with $\epsilon_0 = 2t\sin{k_Fa}$ and $\tilde{O}$ are dimensionless.
An important scaling property of Eq.~\eqref{eq:solitonequ} is that $h(rO(rx))\psi(rx)=r\epsilon\psi(rx)$ follows from
$h(O(x))\psi(x)=\epsilon\psi(x)$, 
i.e., another state $\psi(rx)$ is an eigenstate of $h(rO(rx))$ with energy $r\epsilon$ where $r$ is a dimensionaless scaling factor. This scaling property immediately suggests that it suffices to consider only the following dimensionless Hamiltonian
\begin{equation}\label{eq:solitonequ1}
    h = -\ii \partial_{x} s_z + ( O(\frac{x}{\lambda}) s_+ + \mathrm{h.c.})
\end{equation}
where we drop all tildes for brevity, $r=1/(2\tilde{\alpha}\tilde{m}_0)$, and $\lambda$ is now measured in units of $\lambda^*=ra=a/(2\tilde{\alpha}\tilde{m}_0)$.
Note that we also drop the imaginary factor of $O$ simply because a global phase does not affect the result. The typical dimensionless value $\lambda=1$ therefore corresponds to the dimensionful $\lambda=\lambda^* \approx 20a,60a$ for $\xi/a \approx 3,9$\mycomment{ according to Table.~\ref{Table:elphparamt}}.

\section{Spectral function and optical conductivity}\label{App:Kubo}

At each time slice, we diagnolize the instantaneous electronic Hamiltonian $S^\dag \mathcal{H}_\mathrm{el} S=\mathrm{diag}(\varepsilon_1,\varepsilon_2,\cdots,\varepsilon_N)$. The retarded Green's function in the real space is given by 
\begin{equation}
\begin{split}
    \mathcal{G}^\mathrm{r}(x,y,\omega) 
    &= -\ii \int_{-\infty}^\infty \dd t \ee^{\ii\omega t}\theta(t)\braket{\{ c_x(t),c_y^\dag(0) \}} \\
    &= \sum_l \frac{S_{xl}S^*_{yl}}{\omega-\varepsilon_l+\ii\eta}    
\end{split}
\end{equation}
The spectral function reads $\mathcal{A}(x,y,\omega)=-\frac{1}{\pi}\Im\mathcal{G}^\mathrm{r}(x,y,\omega)$ and we will mainly focus on its diagonal components $\mathcal{A}(x,\omega)=\mathcal{A}(x,x,\omega)$ that reflects the spectral weight on a certain lattice site $x$. 
For our finite-size system one has to set $\eta$ larger than or comparable to the smallest level splitting, which is estimated to be $4t/2N\approx0.001t$, otherwise the finite-size structure becomes visible in the spectral function and the optical conductivity.


The retarded current-current correlator and its spectral expansion is given by 
\begin{equation}\label{eq:jj}
\begin{split}
    \Pi_{\mu\nu}(\omega) 
    &= \frac{-\ii}{\hbar V} \int _0^{\infty} \dd t \ee^{\ii\omega t} \braket{ [j_{\mu}(t), j_{\nu}(0)] } \\
    &= \frac{-\ii}{\hbar V} \int _0^{\infty} \dd t \ee^{\ii\omega t} \,\mathrm{Tr} \left(\hat\rho \,[j_{\mu}(t), j_{\nu}(0)] \right)  \\
    &= \frac{1}{V}\sum_{mn} \frac{\ee^{-\beta \varepsilon_n}-\ee^{-\beta\varepsilon_m}}{Z} \frac{\langle n \lvert j_{\mu}\rvert m\rangle  \langle m \lvert j_{\nu}\rvert n\rangle}{\hbar(\omega +\ii\eta)+\varepsilon_n-\varepsilon_m},
\end{split}
\end{equation}
where we use eigenstates $\mathcal{H}_\mathrm{el} \rvert n\rangle = \varepsilon_n \rvert n\rangle$, the partition function $Z=\mathrm{Tr}(\ee^{-\beta \mathcal{H}})$, the density matrix $\hat\rho=\ee^{-\beta \mathcal{H}}/Z$, and the time dependence of the current operator $\langle n \lvert j_{\mu}(t) \rvert m\rangle 
= \langle n \lvert \ee^{\ii \mathcal{H} t} j_{\mu} \ee^{-\ii \mathcal{H} t} \rvert m\rangle  
=  \ee^{\ii (\varepsilon_n - \varepsilon_m) t} \langle n \lvert j_{\mu}\rvert m\rangle$.
Then, we have the conductivity tensor
\begin{equation}\label{eq:sigma}
\begin{split}
    \sigma _{\mu\nu}(\omega) 
    &= \frac{\ii}{\omega} (\Pi_{\mu\nu}(\omega) - \Pi_{\mu\nu}(0)) \\
    \mycomment{&= -\frac{\Pi_{\mu\nu}(0)}{\omega} + \frac{1}{\hbar \omega V} \int _0^{\infty} \dd t \ee^{\ii\omega t} \braket{ [j_{\mu}(t), j_{\nu}(0)] } \\ } 
    \mycomment{=& \frac{\Pi_{\mu\nu}(0)}{\ii\,\omega} + \frac{\ii}{\omega V}\sum_{mn} \frac{\ee^{-\beta\varepsilon_n}-\ee^{-\beta\varepsilon_m}}{Z} \frac{\langle n \lvert j_{\mu}\rvert m\rangle  \langle m \lvert j_{\nu}\rvert n\rangle}{\hbar(\omega +\ii\eta)+\varepsilon_n-\varepsilon_m} \\}
    &= \frac{\Pi_{\mu\nu}(0)}{\ii\,\omega} + \frac{\ii\hbar}{V}\sum_{mn} \frac{\ee^{-\beta\varepsilon_n}-\ee^{-\beta\varepsilon_m}}{Z} \times\\
    &\frac{\langle n \lvert j_{\mu}\rvert m\rangle  \langle m \lvert j_{\nu}\rvert n\rangle}{\hbar\,\ii\eta+\varepsilon_n-\varepsilon_m}(\frac{1}{\hbar\omega} - \frac{1}{\hbar(\omega +\ii\eta)+\varepsilon_n-\varepsilon_m}) \\
    &= \frac{\ii\hbar}{V}\sum_{mn} \frac{\ee^{-\beta\varepsilon_n}-\ee^{-\beta\varepsilon_m}}{Z (\varepsilon_m-\varepsilon_n)} \frac{\langle n \lvert j_{\mu}\rvert m\rangle  \langle m \lvert j_{\nu}\rvert n\rangle}{\hbar(\omega +\ii\eta)+\varepsilon_n-\varepsilon_m} \\
    &= \frac{\ii\hbar}{V}\sum_{mn} \frac{f_n-f_m}{\varepsilon_m-\varepsilon_n} \frac{\langle n \lvert j_{\mu}\rvert m\rangle  \langle m \lvert j_{\nu}\rvert n\rangle}{\hbar(\omega +\ii\eta)+\varepsilon_n-\varepsilon_m}
\end{split}
\end{equation}
where in the last line we identify the Fermi distribution $n_F(\varepsilon_n)=f_n=\frac{\ee^{-\beta\varepsilon_n}}{Z}$ for this effectively noninteracting system. In the second line, writing the diamagnetic term $\frac{\Pi_{\mu\nu}(0)}{\ii\,\omega}$ with the expansion Eq.~\eqref{eq:jj}, it exactly cancels the first term in the parentheses. 
The electric current operator $j=e v$ assumes velocity operator $v=\frac{\ii}{\hbar}[\mathcal{H},x]$ and
\begin{equation}\label{eq:vel_mat}
    v_{m n}=\langle m \lvert \frac{\ii}{\hbar}[\mathcal{H},x]\rvert n\rangle =\frac{\ii}{\hbar}(\varepsilon_m-\varepsilon_n) x_{m n}.
\end{equation}
Eq.~\eqref{eq:sigma} in the 1D case then gives 
\begin{equation}\label{eq:sigma1}
\begin{split}
    \sigma(\omega)
    &= \frac{\ii\hbar}{V} \sum_{mn} \frac{f_n-f_m}{\varepsilon_m-\varepsilon_n}  \frac{|j_{m n}|^2}{\hbar(\omega +\ii\eta)+\varepsilon_n-\varepsilon_m}\\
    &= -\frac{\ii e^2}{\hbar V} \sum_{mn} \frac{(f_n-f_m)(\varepsilon_n-\varepsilon_m) |x_{m n}|^2}{\hbar(\omega +\ii\eta)+\varepsilon_n-\varepsilon_m},    
\end{split}
\end{equation}
which is our final formula of the optical conductivity.
From the last line of Eq.~\eqref{eq:sigma}, we have the symmetry property for the general conductivity tensor $\sigma_{\mu\nu}(\omega)=\sigma_{\mu\nu}^*(-\omega)$ and hence
\begin{equation}
    \Re\sigma_{\mu\nu}(\omega)=\Re\sigma_{\mu\nu}(-\omega),\Im\sigma_{\mu\nu}(\omega)=-\Im\sigma_{\mu\nu}(-\omega),
\end{equation}
which certainly holds for the longitudinal conductivity $\sigma(\omega)=\sigma_{\mu \mu }(\omega)$ in Eq.~\eqref{eq:sigma1} as well.
Therefore, with the symmetry $\sigma(\varepsilon)=\sigma^*(-\varepsilon)$, we consider only $\varepsilon\geq0$ in Figs.~\ref{Fig:spec_cond}(a1-2) in the main text.

\begin{acknowledgments}
We thank the Max Planck-UBC-UTokyo Center for Quantum Materials for fruitful collaborations and financial support. This work was supported by CFREF, NSERC and CIfAR, JSPS KAKENHI (No.~18H03676), and JST CREST (Nos.~JPMJCR16F1 \& JPMJCR1874). X.-X.Z was also partially supported by the Riken Special Postdoctoral Researcher Program.
\end{acknowledgments}\mycomment{\Yinyang}


\bibliography{reference.bib}  

\end{document}